\documentclass[aps,prb,twocolumn,groupedaddress,floatfix]{revtex4-2}

\usepackage{graphicx}
\usepackage{dcolumn}
\usepackage{bm}
\usepackage{mathtools}
\usepackage{xcolor}
\usepackage[normalem]{ulem}

\def\hc#1{\textcolor{black}{{#1}}}
\usepackage{physics}

\newlength\mylen  \setcitestyle{citesep={,\kern-\mylen}}

\begin{document}

\title{
\hc{Magnetic-field-induced non-adiabatic spin-relaxation resonances in displaced silicon double quantum dots }}
\author{
Sanjay Prabhakar$^{1,2a}$, Shambhu Ghimire$^2$, Roderick Melnik$^3$, Himadri S. Chakraborty$^1$
}
\affiliation{
$^1$\hc{School} of Natural Sciences, Dean L. Hubbard Center for Innovation, Northwest Missouri State University, 800 University Drive, Maryville, MO 64468\\
$^2$Stanford PULSE Institute, SLAC National Accelerator Laboratory, Menlo Park, California 94025, USA\\
$^3$MS2Discovery Interdisciplinary Research Institute, Wilfrid Laurier University, Waterloo, ON, N2L 3C5 Canada \\
$^{a)}$Author to whom correspondence should be addressed: sanjay@nwmissouri.edu, sprabhak@stanford.edu }

\date{Nov 07, 2026}

\begin{abstract}
\hc{Spin relaxation in silicon quantum dots is strongly influenced by spin-orbit coupling, orbital confinement, and magnetic-field-induced level crossings. Here, we investigate acoustic-phonon-mediated spin relaxation in laterally displaced silicon double quantum dots under combined in-plane and out-of-plane magnetic fields. Using finite-element diagonalization of a spin-orbit-coupled effective-mass Hamiltonian, we show that spatial displacement modifies the orbital spectrum and produces additional spin-relaxation resonances, the hot spots, beyond those of a single dot. The resulting relaxation landscape exhibits strongly varying lifetimes, including millisecond-scale regimes from higher excited states. We identify the interplay between the dot separation and magnetic field strength as a mechanism governing the emergence of these non-adiabatic features. Our results establish spatial displacement as a control parameter for engineering spin relaxation in silicon double quantum dots and suggest a route toward tunable spin-lifetime conditions in semiconductor spin-qubit devices.}
\end{abstract}

\maketitle
\section{Introduction}

Gate-controlled manipulation of single and double electron spins in quantum dots (QDs) using electric and magnetic fields is a key ongoing effort toward solid-state quantum computing and quantum information processing~\cite{sobrino24,khomitsky25,bhat22,suarez25,banerjee25,raith12,raith11,bulaev07,sp24,zwerver23,paquelet23}. Single and double QDs, created using advanced techniques like molecular beam epitaxy and metal oxide chemical vapor deposition  with lithographic processes in metal oxide field effect transistors, are crucial for qubit gate operations. This enables initialization of one \hc{qubit} or two qubits from source to drain~\cite{wang22,fernandez22,zhou23,jock18,burkard23,steinacker25,woods23}. Advanced \hc{technologies allow} rapid spin-current measurements in silicon QDs, which are preferred over GaAs due to \hc{their} small spin-orbit coupling effects resulting in longer qubit coherence times for more effective gate operations~\cite{gilbert23,liles24,jock18}. In the \hc{laboratory}, long spin-relaxation ($T_1$) and decoherence ($T_2\approx 2T_1$) times can be achieved by mechanisms like electron-phonon (non-adiabatic) interaction, electron-hyperfine interactions in the presence of spin-orbit coupling. \hc{This enables} qubit spins to be reliably initialized, manipulated, and read out which is a key requirement for quantum computing~\cite{wang24,appel22,haldar25,kedim25,dyte25,kazemi26,zhang26,memon24,oakes23}.

In silicon QDs, weak spin-orbit coupling is incorporated into the Hamiltonian via Rashba and linear Dresselhaus terms~\cite{raith12,jock18,debnath24,ferreira23,hetenyi22}. The Rashba spin-orbit coupling originates from structural inversion asymmetry along the growth direction, while the Dresselhaus spin-orbit coupling arises from bulk inversion asymmetry of the crystal lattice. Spin relaxation from electron–phonon interactions (piezo and acoustic) in QDs creates spin-hot spots through level crossings between spin and orbital states~\cite{raith12,sp24,bulaev07,wang24,hosseinkhani22}. In silicon QDs, only acoustic phonons significantly contribute to spin flips, unlike in III-V semiconductors where piezo-phonons are strong~\cite{kolok25,coste23,han22,sp13,sp24}. Spin relaxation in silicon and germanium \hc{single} dots has been extensively studied~\cite{hsueh23,bosco22,hosseinkhani22,liu22,blumoff22,huang18}, \hc{while that in silicon double dots} has recently been observed experimentally \hc{with prospects in} quantum computing applications~\cite{lundberg20,ciriano21,crepieux22,ginzel23,hao14}. Experiments show that in silicon double dots, two qubit spins can be initialized in s and p states and be driven into superposition at spin-hot spots using rapid adiabatic magnetic field pulses~\cite{zhao22,krzywda20,takeda24}. These theoretical and experimental studies reported results on generating two-qubit superposition states for quantum computing and quantum information processing.

\hc{Spin-relaxation resonances or spin-hot spots} arise from strong mixing of singlet and triplet spin and orbital states, leading to enhanced phonon-mediated spin relaxation in admixed qubit spin states, which has been widely studied~\cite{jock22,spence22,sp13,sp24,raith12,bulaev07}. In Ref.~\cite{raith11}, spin-hot spots are \hc{predicted} in \hc{displaced} double QDs, while Ref.~\cite{bulaev07} observes two spin-hot spots of heavy holes in a single dot when in-plane and out-of-plane magnetic fields are applied. The current study identifies a new source of non-adiabatic spin–hot-spot mechanism that emerges when in-plane and out-of-plane magnetic fields are applied as the two \hc{silicon} dots are pulled apart from their origin.
\hc{This may include hot spots with spin-relaxation rates separated by several orders of magnitude driven by the magnetic-field variation, although there may exist {\em cold} sectors of the field that do not support any hot spot.}
\hc{With increasing inter-dot separation, new hot spots may emerge in the lower magnetic field region due to strong enough confinement potential from combined in-plane and out-of-plane magnetic fields. For a well displaced dot pair, even changes in the out-of-plane magnetic field strength may significantly alter the hot spot structure and rate as a function of the in-plane field.} In Ref.\,\cite{jock22,jock18}, the authors propose preparing two qubits in a singlet state and reading out superposition states, exploiting singlet–triplet overlap induced by non-adiabatic tuning of the detuning parameter (magnetic field). Hence, spin-hot spots provide ideal qubit readout states, where preserving superposition information critically depends on decoherence times $T_2\le 2T_1$. This paper demonstrates the existence of multiple spin-hot spots, some with decoherence times up to four to six orders of magnitude longer than others, making them ideal for qubit manipulation in quantum information processing.

The paper is organized as follows. Section \ref{theoretical-model} presents a theoretical model describing spin relaxation in single and double silicon QDs arising from interactions with acoustic phonons. Section \ref{computational-method} briefly outlines a computational diagonalization approach using finite element method simulations to determine the energy spectrum and evaluate the matrix elements governing phonon-mediated spin relaxation rates.
Section \ref{results-and-discussions} presents and analyzes results of electron spin relaxation in \hc{double QDs, driven by acoustic phonons with the combinations of inter-dot separation and both in-plane and out-of-plane magnetic fields being the control parameters. Experimental  ideas to test the results are discussed in Section \ref{experimentalA}.} Finally, Section \ref{conclusion} provides a summary of the main results.

\section{Theoretical Model}\label{theoretical-model}

The study models an electron’s three-dimensional motion in a quantum dot under both in-plane and out-of-plane magnetic fields, incorporating the Rashba and linear Dresselhaus spin–orbit couplings. Following established approaches from prior studies~\cite{corrigan23,raith11,raith12,bulaev07,sp13,sp24}, the total Hamiltonian can be written as
\begin{equation}
H = H_0 + H_1 + H_R + H_D, \label{H}
\end{equation}
\begin{eqnarray}
H_0 &=& \frac{P_x^2}{2m_x} + \frac{P_y^2}{2m_y}+\frac{P_z^2}{2m_z} + \frac{1}{2}m \omega_0^2 V(x,y) +eEz, ~~~ \label{H0} \\
 H_1 &=& \frac{1}{2}\mu_B \left(g_x B_x \sigma_x + g_y B_y \sigma_y + g_z B_z \sigma_z \right), ~~~ \label{Hz} \\
 H_R &=& \alpha_R\left(\sigma_xP_y - \sigma_y P_x\right),\label{HR}\\
  H_D &=& \alpha_D\left(-\sigma_xP_x + \sigma_yP_y\right).\label{HD}
\end{eqnarray}
Here, $\mathbf{P}=\mathbf{p} + e \mathbf{A}$ with the vector potential, $\mathbf{A}= \left(-y B_z/2, xB_z/2, yB_x-xB_y\right)$ and $\mathbf{p}$ is the momentum operator. $V_z=Ez$, \hc{$E$ being the electric field strength,} is the confinement potential in the \hc{out-of-plane} z-direction due to band offset of semiconductor material at the interface of the heterojunction. $V(x,y)$ is the confinement potential in the \hc{in-plane} (lateral) direction.   $H_1$ is the Zeeman spin-splitting energy due to applied in-plane magnetic field strength \hc{$B_{\mbox{\scriptsize in}}$ ($B_x$ and $B_y$ combined)} and out-of-plane magnetic field strength $B_z$. Here, $m_i (i=x,y,z)$ is the effective mass of electron in the QD, $\ell_0 = \sqrt{\hbar/m\omega_0}$ is the radius of the lateral size of a dot and $g_i (i=x,y,z)$ is the bulk g-factor of electron. The spin-orbit coupling, $H_{so} = H_R+H_D$, where $H_R$ and $H_D$ are, respectively, the Rashba and the linear Dresselhaus spin-orbit couplings. Also, $\sigma_x$, $\sigma_y$ and $\sigma_z$ are the Pauli spin matrices. The lateral confinement potential, $V_{xy}$, of the single and double dot can be written as
\begin{eqnarray}
V(x,y) &=& x^2+y^2, \mbox{single QD} \label{vxy1} \\
V(x,y) &=& x^2+\frac{1}{4y_0^2} (y^2-y_0^2)^2, \mbox{double QDs}. \label{vxy2}
\end{eqnarray}
For double QDs, $y_0$ represents the displacement of the dots from the origin. We focus on evaluating the phonon-induced spin relaxation rate at low temperatures, where acoustic phonons dominate and thermal excitations are minimal. The calculation aims to provide theoretical descriptions of how electron–phonon interacts, together with spin–orbit coupling, lead to spin-flip transitions between QD energy states, thereby determining the spin relaxation time relevant for qubit decoherence. Following Refs.~\cite{raith11,raith12,prada08}, the interaction between electron and acoustic phonon can be written as
\begin{equation}
u^{\mathbf{q}\lambda}_{ph}\left(\mathbf{r}\right)=i \sum_{\lambda=l,t} \sqrt{\frac{\hbar}{2\rho V \omega_{\mathbf{q}\lambda}}} e^{i \mathbf{q\cdot r} } q D_q^\lambda b^{\dag}_{\mathbf{q}\lambda} + H.c.,
\label{u}
\end{equation}
\begin{equation}
D_q^\lambda= \Xi_d \mathbf{\hat{e}_q^\lambda \cdot \hat{q}} + \Xi_u \hat{e}_{q_z}^\lambda  \hat{q_z},
\label{Dq}
\end{equation}
where $\rho$ is the crystal mass density, $V$ is the volume of the dot.  Also, $b^{\dag}_{\mathbf{q}\alpha}$ creates an acoustic phonon with wave vector $\mathbf{q}$ and polarization $\mathbf{\hat{e}_q^\lambda}$, where $\lambda=l,t_1,t_2$ are chosen as one longitudinal and two transverse phonon modes. The polarization directions of the induced phonon are $\mathbf{\hat{e}_q}^l=\left(\sin\theta \cos\phi, \sin\theta \sin\phi, \cos\theta \right)$,   $\mathbf{\hat{e}_q}^{t_1}=\left(-\sin\phi, \cos\phi, 0 \right)$, and $\mathbf{\hat{e}_q}^{t_2}=\left(\cos\theta \cos\phi, \cos\theta \sin\phi, -\sin\theta \right)$. The deformation potential strength is determined by the dilatation potential constant, $\Xi_d$, and shear potential constant, $\Xi_u$.  Based on the Fermi Golden Rule, the phonon induced spin transition rate is given by~\cite{raith11,raith12,prada08}
\begin{equation}
\Gamma=\frac{2\pi}{\hbar}\int \frac{d^3\mathbf{q}}{\left(2\pi\right)^3}\sum_{\lambda=l,t}\arrowvert M\left(\mathbf{q}\alpha\right)\arrowvert^2\delta\left(\hbar s_\lambda \mathbf{q}-\varepsilon_{f}+\varepsilon_{i}\right),
\label{1-T1}
\end{equation}
where  $s_l$,$s_t$ are the longitudinal and transverse acoustic phonon velocities in Si QDs.  The matrix element $M\left(\mathbf{q}\lambda\right)=\langle \psi_i|u^{\mathbf{q}\lambda}_{ph}\left(\mathbf{r},t\right)|\psi_f\rangle$ with the emission of one phonon is calculated numerically~\cite{comsol}. Here $|\psi_i\rangle$ and $|\psi_f\rangle$ correspond to the initial and finial states of the Hamiltonian $H$. More precisely, we write the spin-relaxation rate due to longitudinal and transverse phonon as
\begin{widetext}
\begin{eqnarray}
\Gamma^{l}&=& \frac{\Delta E^5} {210\pi\rho\hbar^6 s_l^7}
\begin{bmatrix}
\left(35 \Xi_d^2 + 3 \Xi_u^2 + 14 \Xi_u \Xi_d \right) \left(|M_x|^2 + |M_y|^2\right)
 + \left(35 \Xi_d^2 + 15 \Xi_u^2 + 42 \Xi_u \Xi_d \right)|M_z|^2
\end{bmatrix},
\label{W01}\\
~\nonumber\\
\Gamma^{t}&=& \frac{\Delta E^5 \Xi_u^2} {105\pi\rho\hbar^6 s_t^7}
\begin{bmatrix}
2\left(|M_x|^2 + |M_y|^2\right) + 3 |M_z|^2
\end{bmatrix}.
\label{W0t}
\end{eqnarray}
\end{widetext}
where $M_x = \langle \psi_i|x|\psi_f\rangle$, $M_y=\langle \psi_i|y|\psi_f\rangle$, and $M_z=\langle \psi_i|z|\psi_f\rangle$.
\begin{figure}
\includegraphics[width=8.5cm,height=6cm]{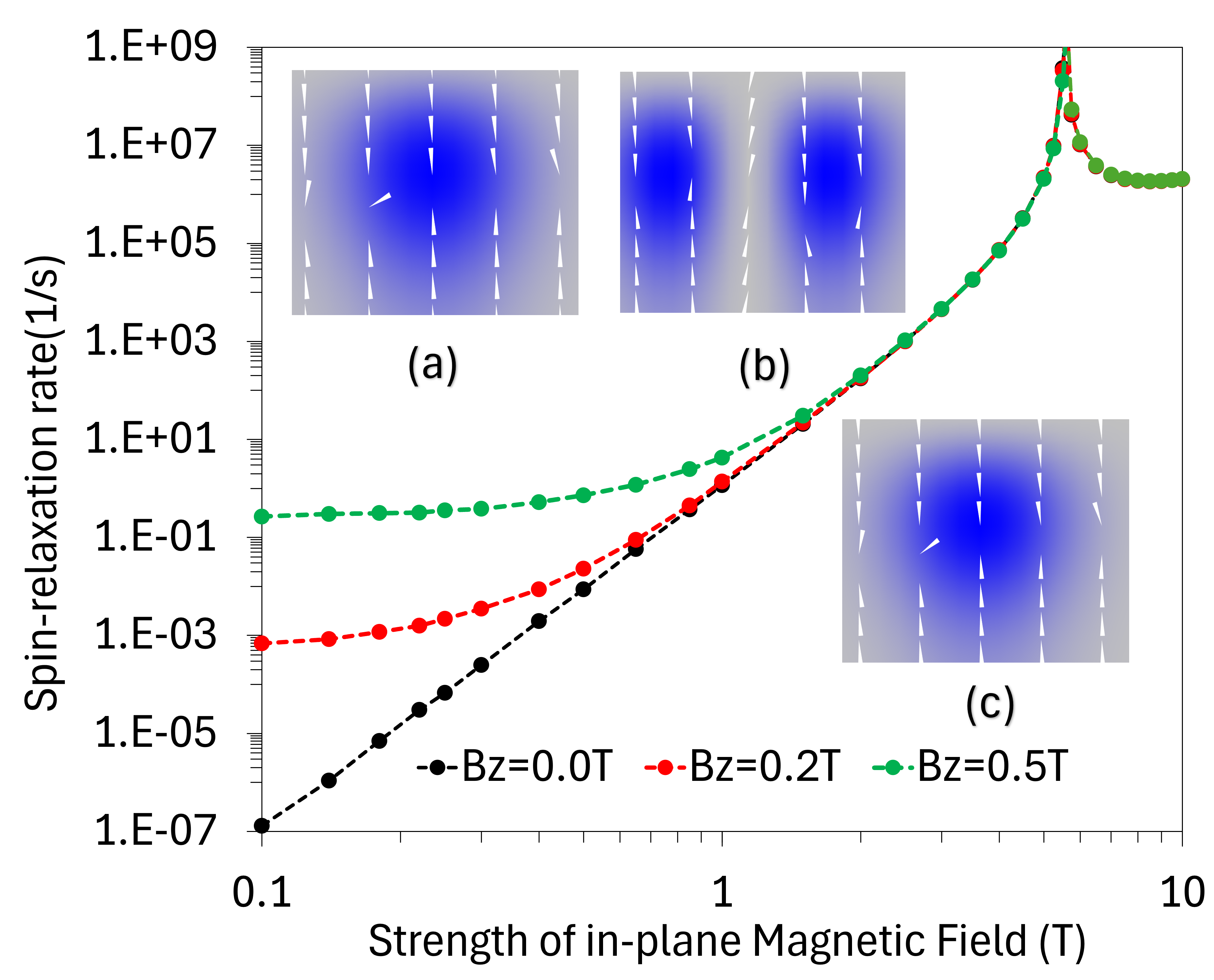}
\caption{
\label{fig1} (Color online) Spin-relaxation rate versus the strength $B_{\mbox{\scriptsize in}}$ of in-plane magnetic field in a single quantum dot in the presence of out-of-plane magnetic fields at $B_z$ = 0 T (black), 0.2 T (red), and 0.5 T (green). Spin-hot spots can be seen in the vicinity of $B_{\mbox{\scriptsize in}}$ = 5.5 T.
Insets (a), (b) and (c) are the probability density of the three lowest electron spin states for \hc{$B_{\mbox{\scriptsize in}}$ = 5.75 T and $B_z$ = 0.2 T, close to the hot spot. Insets also show} the kinetic momentum density (white surface arrows). We chose $\ell_0$ = 25 nm.
}
\end{figure}

\section{Computational Method}\label{computational-method}

We suppose that a  QD is formed at the center of a $1200\times 1200~\mathrm{nm^2}$ geometry.  We then diagonalize the total Hamiltonian $H$ numerically using the Finite Element Method~\cite{comsol}.  Since the geometry is much larger compared to the actual lateral size of the QD, we impose Dirichlet boundary conditions, find the eigenvalues, eigenfunctions and the matrix elements $M\left(\mathbf{q}\lambda\right)$ of $H$. The results presented in Fig.\,\ref{fig1} is consistent with the Fig.\,5 of Ref.\,\cite{raith11} for a single QD, \hc{even though} with a slight modification in the model \hc{where} we consider three-dimensional QDs formed at the heterojunction of Si-SiO$_2$ layer. The material constants of silicon for the simulations are taken as follows.  We use the parameters of a silicon QD grown along the [001] direction with silicon dioxide on the top. The two-dimensional electron gas is defined at the heterojunction of Si-SiO$_2$ layer. The effective masses are $m_{xy}=0.19$ and $m_z=0.98$ for silicon and $m_{ox}=0.42$ for silicon-oxide layer. The bulk g-factor is chosen to be $g_0=2$. The electric field strengths that control the Rashba spin-orbit coupling is 3.1 MV/m and the Dresselhaus spin-orbit coupling is 1.0 MV/m. The choice of these makes the Dresselhaus spin-orbit coupling strength about three times larger than the Rashba spin-orbit coupling, which is consistent to the experimental results in Ref.~\cite{jock18}. To calculate the spin-relaxation due to acoustic phonon, we use $s_l$ = 9150 m/s for LA phonons, $s_t$ = 5000 m/s for TA phonons, and $\rho$ = 2330 kg/m$^3$. The choice of deformation potential constants is consistent with Ref.\cite{raith11} as $\Xi_d$ = 5 eV and $\Xi_u$ =9 eV.

\section{Results and Discussions}\label{results-and-discussions}

\hc{\subsection{Variation of inter-dot separation}}

In Fig.\,\ref{fig1}, we plot the spin-relaxation rate \hc{as a function of the strength $B_{\mbox{\scriptsize in}}$ of in-plane magnetic field in the presence of three selections of relatively weak out-of-plane magnetic field, $B_z$ = 0 T, 0.2 T, and 0.5 T, for a single QD of size $\ell_0$ = 25 nm. A spin-hot spot is observed around $B_{\mbox{\scriptsize in}}$ = 5.5 T. Note that the spin relaxation rate is rather insensitive to a choice of B$_z$ over the range $B_{\mbox{\scriptsize in}}>$ 1.5 T. When $B_{\mbox{\scriptsize in}}<$ 1.5 T, on the other hand, the relaxation rates differ significantly among the values of B$_z$. We plot in the insets, Fig.\,\ref{fig1} (a), (b) and (c), the probability density of three lowest electron spin states in the close proximity ($B_{\mbox{\scriptsize in}}$ = 5.75 T, $B_z$ = 0.2 T) of the spin-hot spot. The inset images further display the kinetic momentum density by white surface arrows. The level crossing between the spin and orbital states is clearly observed in (b). We have verified that for $B_z$ = 0 T, the spin-relaxation rate vanishes as the seventh-power of $B_{\mbox{\scriptsize in}}$ with the field approaching zero, which was also seen in Ref.\,\cite{raith12}. However, there is no such clear pattern of the relaxation rate decrease versus $B_{\mbox{\scriptsize in}}$ in the presence of $B_z$, indicating complicated mathematical structure of phonon interaction with non-zero $B_z$.}
\begin{figure*}
\includegraphics[width=17cm,height=7cm]{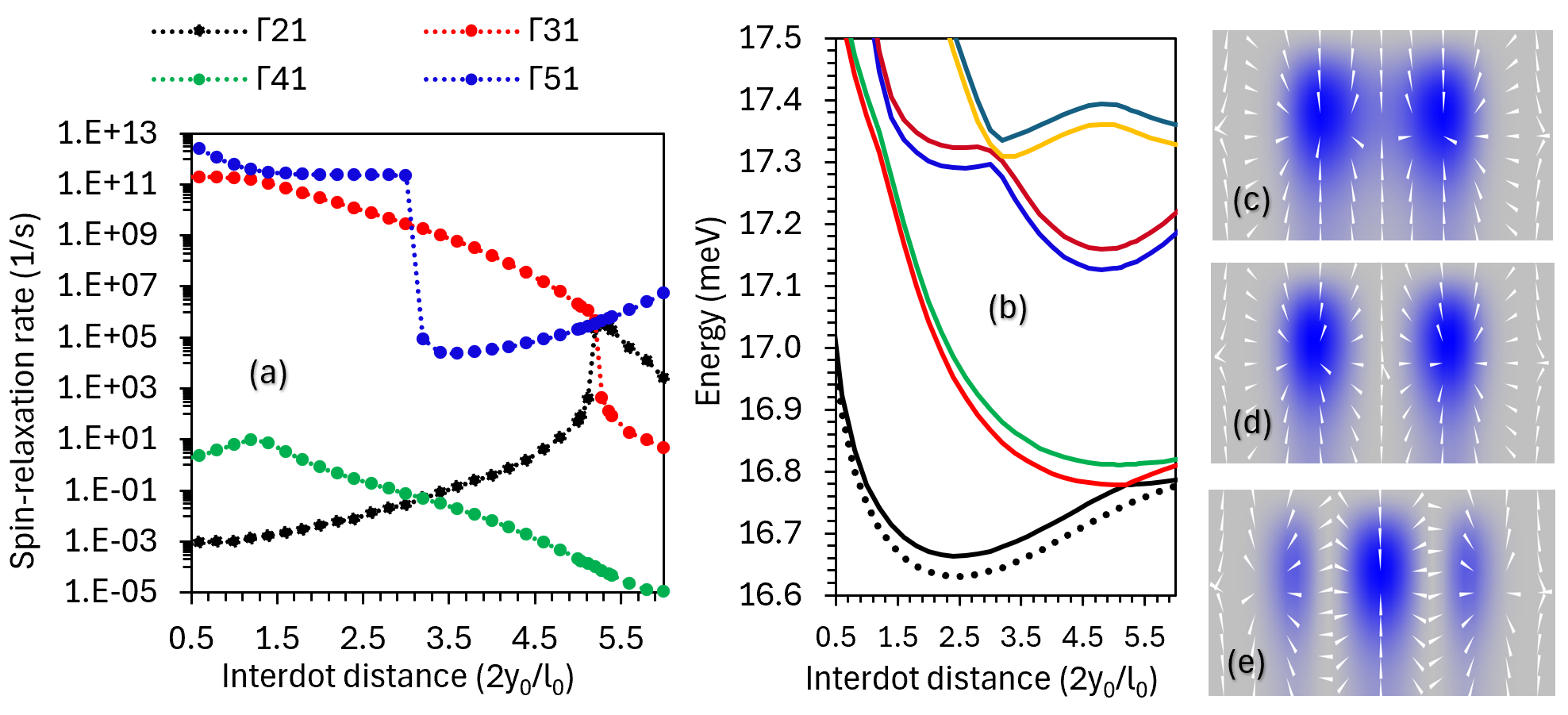}
\caption{
\label{fig2} (Color online) (a) \hc{Spin-relaxation rate versus inter-dot distances, in the unit of a dot size $\ell_0$ = 25 nm, for a double quantum dot, each located at $\pm y_0$ from their midpoint (origin), for four color-identified rates $\Gamma N1$ to the ground state (1). A spin-hot spot is seen when the dots are center-to-center displaced with $y_0$ = 66 nm due to the avoided crossing between the black and red bands seen in (b). The probability densities of ground, 3rd and 5th spin states near this hot spot, along with the kinetic momentum density (white surface arrows), are shown, respectively, in (c), (d), and (e). We chose $B_z$ = $B_{\mbox{\scriptsize in}}$ = 0.2 T. In (c) to (e), we chose $y_0$ = 60 nm.}
}
\end{figure*}

\hc{Fig.~\ref{fig2} shows rates $\Gamma N1$ of four spin-relaxation from the excited states $N$ = 2--5 to the ground state (1) as the inter-dot distance of a double QD varies in a weak magnetic field of $B_z$ = $B_{\mbox{\scriptsize in}}$ = 0.2 T. The distance is plotted in the unit of the single dot size $\ell_0$ = 25 nm where $2y_0$ is the center-to-center separation of the dots with the origin of the coordinate system is the midpoint between the dots. Obviously, $y_0$ is a parameter that controls the separation of the dots along $y$-direction from the system's center. An unusual non-adiabatic spin-hot spot, previously unseen, is predicted closely around $2y_0/\ell_0$ = 5.28 or $y_0$ = 66 nm. This hot-spot arises due to an avoided crossing between the second (black) and third (red) excited state bands seen in panel (b). The green curve for the 4-to-1 spin-relaxation suggests that the rate decreases, which enhances the decoherence time for potential qubit gate operation. Here obviously, opposite spin-states between the left and right QD are making transitions. In Figs.\,\ref{fig2} (c), (d), and (e), we respectively plot the probability densities of the ground, third and fifth spin states near this hot-spot at $y_0$ = 60 nm, also showing the kinetic momentum density surface arrows. Probability densities of second and fourth states look similar to (c) and (d) with slight variations due to spin-splitting behavior in the presence of a $B$ field.  In the density of ground state in (c), we observe a separation between two dots with some wavefunctions overlap. (c) also suggests a different density than in (d) and (e) due to the inter-dot separation driven avoided crossing of the bands mentioned above and displayed in (b).}
\begin{figure*}
\includegraphics[width=18cm,height=7cm]{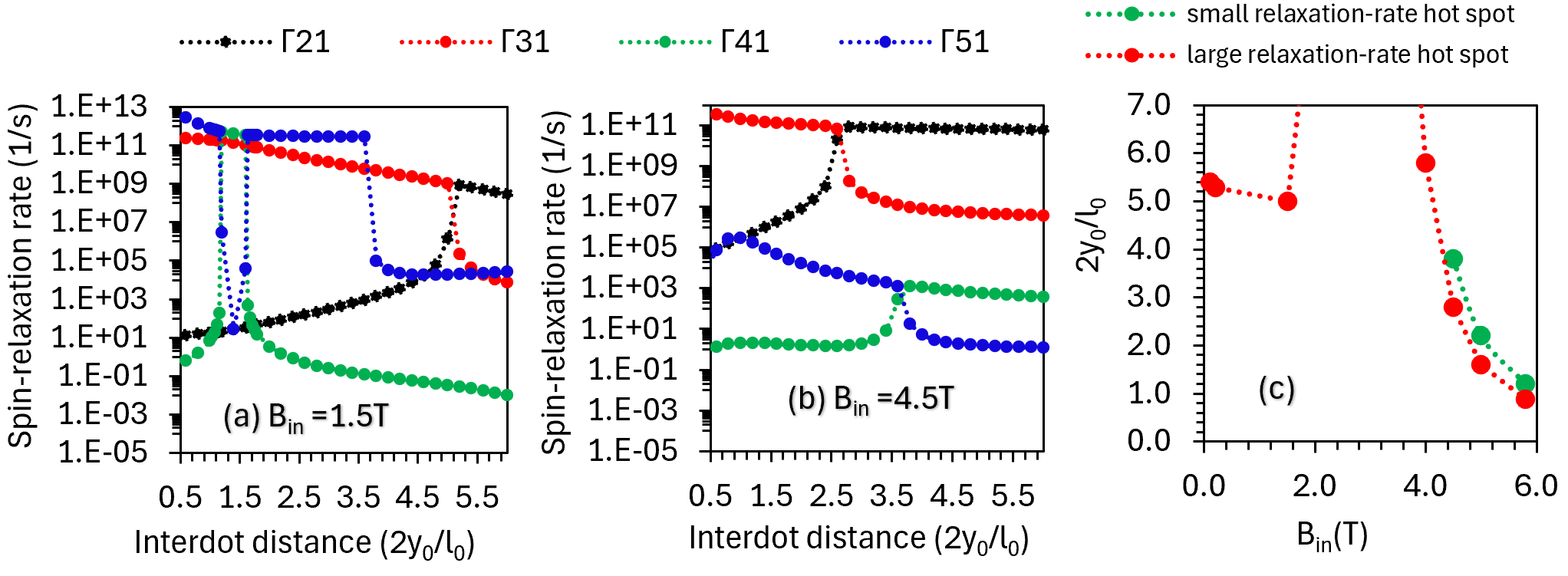}
\caption{
\label{fig2b} (Color online) \hc{Spin-relaxation rates $\Gamma N1$ versus inter-dot distances for $B_{\mbox{\scriptsize in}}$ = 1.5 T (a) and $B_{\mbox{\scriptsize in}}$ = 4.5 T (b). As in Figs.\ref{fig2} and \ref{fig3}, we kept $B_z$ = 0.2 T. In (c) are plotted the hot spot emerging distances with $B_{\mbox{\scriptsize in}}$ between $\Gamma 21$ and $\Gamma 31$ (red) as well as between $\Gamma{41}$ and $\Gamma{51}$ (green). Hot spots do not appear over a range around $B_{\mbox{\scriptsize in}}$ = 1.8-3.8 T for the former.}
}
\end{figure*}

\hc{Fig.~\ref{fig2b} presents the same four spin-relaxation rates versus the inter-dot distance for stronger $B_{\mbox{\scriptsize in}}$ = 1.5 T (a) and $B_{\mbox{\scriptsize in}}$ = 4.5 T (b). Note that, as in Figs.~\ref{fig2}, Fig.~\ref{fig2b} also retains $B_z$ at a weak value of 0.2 T. Results show that stronger $B_{\mbox{\scriptsize in}}$ values considered induce new hot spots from the avoided crossing between $\Gamma 41$ and $\Gamma 51$ with a dramatically reduced hot-spot rate from the other feature. In fact, the corresponding spin relaxation times elongate several orders of magnitude, roughly, from nano to milliseconds! In (c), we provide a general plot of the ideal distances where hot-spots, between $\Gamma{21}$ and $\Gamma{31}$ (red curve) as well as between $\Gamma{41}$ and $\Gamma{51}$ (green curve), emerge as $B_{\mbox{\scriptsize in}}$ varies. As shown at roughly $B_{\mbox{\scriptsize in}}$ = 1.8  to 3.8 T for the former, there appear avoided regimes, or cold sectors, where no hot spot emerges. Because $B_z$ = 0.2 T is kept constant and significantly weak, as in Fig.~\ref{fig2}, and because we found qualitatively similar behavior in the results (not shown) of $B_z$ = 0 T, we conclude that $B_{\mbox{\scriptsize in}}$ is predominantly responsible for inducing hot spot groups of vastly different rates (lifetimes). Panel (c), showing the ideal locations of the spin-hot spot in a graph of dot separation ($2y_0/\ell_0$) versus in-plane field $B_{\mbox{\scriptsize in}}$, suggests a critical interplay among the quasi-Hamiltonian (arising from dots displaced from the origin), the in-plane magnetic field, and, more generally, the out-of-plane magnetic field when large enough (see below).}

\hc{\subsection{A theoretical framework}}

Ref.\,\cite{raith11} has investigated the band avoided-crossing as double QDs are pulled apart from the origin in the presence of zero magnetic field. In Fig.\,\ref{fig2}, we find that splitting of bands are possible and hence able to support the unusual non-adiabatic spin-hot spot when dots are separated at about 132 nm in the presence of both $B_{\mbox{\scriptsize in}}$ and $B_z$. Ref.~\cite{bulaev07} has studied the influence of $B_{\mbox{\scriptsize in}}$ and $B_z$ in single heavy hole quantum dots, though it was not quite clear how $B_{\mbox{\scriptsize in}}$, that has $z$-component of the vector potential, is implemented in the two-dimensional Hamiltonian. Hence, we consider a complete Hamiltonian (\ref{H}) and apply the canonical momentum, $p_z=\hbar k =0$ at $\Gamma$-point to write the two dimensional Hamiltonian as,
\begin{equation}
H = H_{0} + H_2  +  H_R+H_D,
\label{H-1}
\end{equation}
where $B_{\mbox{\scriptsize in}}$ field components ($B_x, B_y$) can be incorporated in $H_2$ and  $g_z > g_{x,y}$ \cite{raith11}. The terms $H_R$ and $H_D$ are same as in Eq.(\ref{H}), while the remaining terms are:
\begin{equation}
H_{0} = \frac{P_x^2+P_y^2}{2m}  + {\frac{1}{2}} m \omega_o^2  \left(r\pm r_0\right)^2 + \frac{\Delta}{2}  \sigma_z,
\label{hxy-1}
\end{equation}
\begin{equation}
H_{2} = \frac{e^2}{2m_z}\left( B_y x - B_x y \right)^2,
\label{H2}
\end{equation}
in which $\Delta=g_z\mu_B B_z$. At this point, we introduce the relative coordinate $\mathbf{R}=\mathbf{r}\pm\mathbf{r_0}=\left(X,Y,0\right)$ and the relative momentum $\mathbf{P}=\mathbf{p}\pm\mathbf{p_0}=\left(P_X,P_Y,0\right)$, where $\mathbf{p_0}=m\mathbf{\dot{r}}_0$ is the momentum of the displaced dot which vanishes due to the fact that the displaced double dots from the origin are stationary. The $\pm$ sign in $p_0$ and $r_0$ represents left and right quantum dots displaced from the origin. Consequently, the Hamiltonian~(\ref{H-1}) can be written in terms of the relative-coordinate and momentum as,
\begin{equation}
H=H_0\left(\mathbf{P},\mathbf{R}\right)+H_{so}\left(\mathbf{P},\mathbf{R}\right)+
H_{ad}\left(\mathbf{P},\mathbf{R};\mathbf{p_0},\mathbf{r_0}\right),
\label{total-adiabatic}
\end{equation}
where
\begin{eqnarray}
H_0\left(\mathbf{P},\mathbf{R}\right) = {\frac {1}{2m}}\left\{\mathbf{P}+e\mathbf{A}(\mathbf{R})\right\}^2 + {\frac{1}{2}} m \omega_o^2R^2 + {\frac \Delta 2}  \sigma_z,\label{tilde-hxy}~~~
\end{eqnarray}
\begin{eqnarray}
 H_R\left(\mathbf{P},\mathbf{R}\right) &=& \alpha_R\left(\sigma_x \left\{P_Y+eA_Y\right\} - \sigma_y \left\{P_X+e A_X\right\}\right),\label{HR-1}~~~\\
  H_D\left(\mathbf{P},\mathbf{R}\right) &=& \alpha_D\left(-\sigma_x  \left\{P_X+e A_X\right\} + \sigma_y \left\{P_Y+eA_Y\right\} \right).~~~\label{HD-1}
\end{eqnarray}
\begin{eqnarray}
H_{ad}\left(\mathbf{P},\mathbf{R};\mathbf{p}_0,\mathbf{r}_0\right)&=&\frac{1}{m}\left\{\mathbf{P}
+e\mathbf{A}\left(\mathbf{R}\right)\right\}\cdot \left\{\mathbf{P}+e\mathbf{A}\left(\mathbf{R}\right)\right\}\nonumber\\
&&+H_{so}\left(\mathbf{p}_0,\mathbf{r}_0\right).\label{Had}
\end{eqnarray}
The  Hamiltonian $H_0\left(\mathbf{P},\mathbf{R}\right)$, $H_R\left(\mathbf{P},\mathbf{R}\right)$, $H_D\left(\mathbf{P},\mathbf{R}\right)$ can be diagonalized on the basis of the number states $|n_+,n_-,\pm 1\rangle$, which is the same for left and right dots:
\begin{equation}
H_0\left(P,R\right)=\left(N_+ + \frac{1}{2}\right)\hbar\Omega_+ + \left(N_- + \frac{1}{2}\right)\hbar\Omega_-  + \frac{\Delta}{2}\sigma_z,\label{H0}
\end{equation}
where $N_{\pm}=a^\dagger_{\pm}a_{\pm}$ are the number operators with eigenvalues $n_\pm \in N_0$. Here,
\begin{eqnarray}
a_{\pm}=\frac{1}{\sqrt{4m\hbar\Omega}}\left(iP_x \pm P_y\right)+\sqrt{\frac{m\Omega}{4\hbar}}\left( X \mp i Y \right),\\ \label{a-lowering}
a_{\pm}^\dagger=\frac{1}{\sqrt{4m\hbar\Omega}}\left(-iP_x \pm P_y\right)+\sqrt{\frac{m\Omega}{4\hbar}}\left( X \pm i Y \right),\label{a-raising}
\end{eqnarray}
provided that $\left[a_{\pm},a_{\pm}^\dagger\right]=1$.
Correspondingly, the other terms may also be expressed in terms of the raising and lowering operators,
\begin{eqnarray}
H_R\left(\mathbf{P},\mathbf{R}\right) &=&\alpha_R\left(\xi_+\sigma_+ a_+ - \xi_- \sigma_- a_-\right)+H.c.,~~~~~~\label{Hso-1}\\
H_D\left(\mathbf{P},\mathbf{R}\right) &=& i\alpha_D\left(\xi_+\sigma_- a_+ + \xi_- \sigma_+ a_-\right)+H.c.,~~~~~~\label{Hso-2}
\end{eqnarray}
where $\xi_{\pm}= \sqrt {m\Omega/\hbar}\pm eB/\sqrt{4m\hbar\Omega}$, $\sigma_{\pm}=\left(\sigma_x\pm i\sigma_y\right) /2$, $\omega_{\pm}=\omega\left( 1 \pm \omega_c/\left(2\omega\right)\right)$, $\Omega_{\pm}=\Omega \pm \omega_c/2$ and $\Omega=\sqrt {\omega_0^2+\omega_c^2/4}$ with $\omega_c=eB/m$  being the cyclotron frequency. In Eqs.~(\ref{Hso-1}) and~(\ref{Hso-2}), $H.c.$ denotes the Hermitian conjugate. For a special case, when $B_x=B_y$, we can also write $H_{ad}$ and $H_2$ in terms of number operator which plays important role on admixture mechanism of spin states as left and right dots are pulled apart.
\begin{widetext}
\begin{eqnarray}
H_{ad}=\frac{\hbar}{2}\left(\xi_+ a_+-\xi_- a_-\right)\omega_+ x_0^{L,R}
+\frac{1}{\hbar}\left(\alpha_R -i\alpha_D \right)m\omega_+\sigma_+ x_0^{L,R} +H.c., \label{had-2}\\
H_2=\frac{e^2B_x^2}{2m_z}\left\{\ell^2\left( a_-^2 + a_-a_+^\dagger + a_+^\dagger a_- + (a_+^\dagger)^2  \right) + 2 \ell \left( a_- + a_+^\dagger\right) x_0^{L,R} + \left(x_0^2\right)^{L,R}  \right\} \label{H2-2},
\end{eqnarray}
\end{widetext}
where $x_0^L > 0$ for the left dot and $x_0^R < 0$ for the right dot. As can be seen in Eq.~\ref{H2-2}, the double-dot Hamiltonian depends on the quadratic function of the separation between the dots. As a result in Fig.\ref{fig2}, we find parabolic bands as a function of inter-dot distances in a double QD. The parabolic bands cross when dots are separated at around 132 nm, which induces a spin-hot spot, evidently seen in  Fig.~\ref{fig2} (a). \hc{The above formalism also serves as a framework to justify the emergence of multiple hot spots when $B$ varies for fixed inter-dot separations, as found in the following results.}
\begin{figure*}
\includegraphics[width=18.0cm,height=7cm]{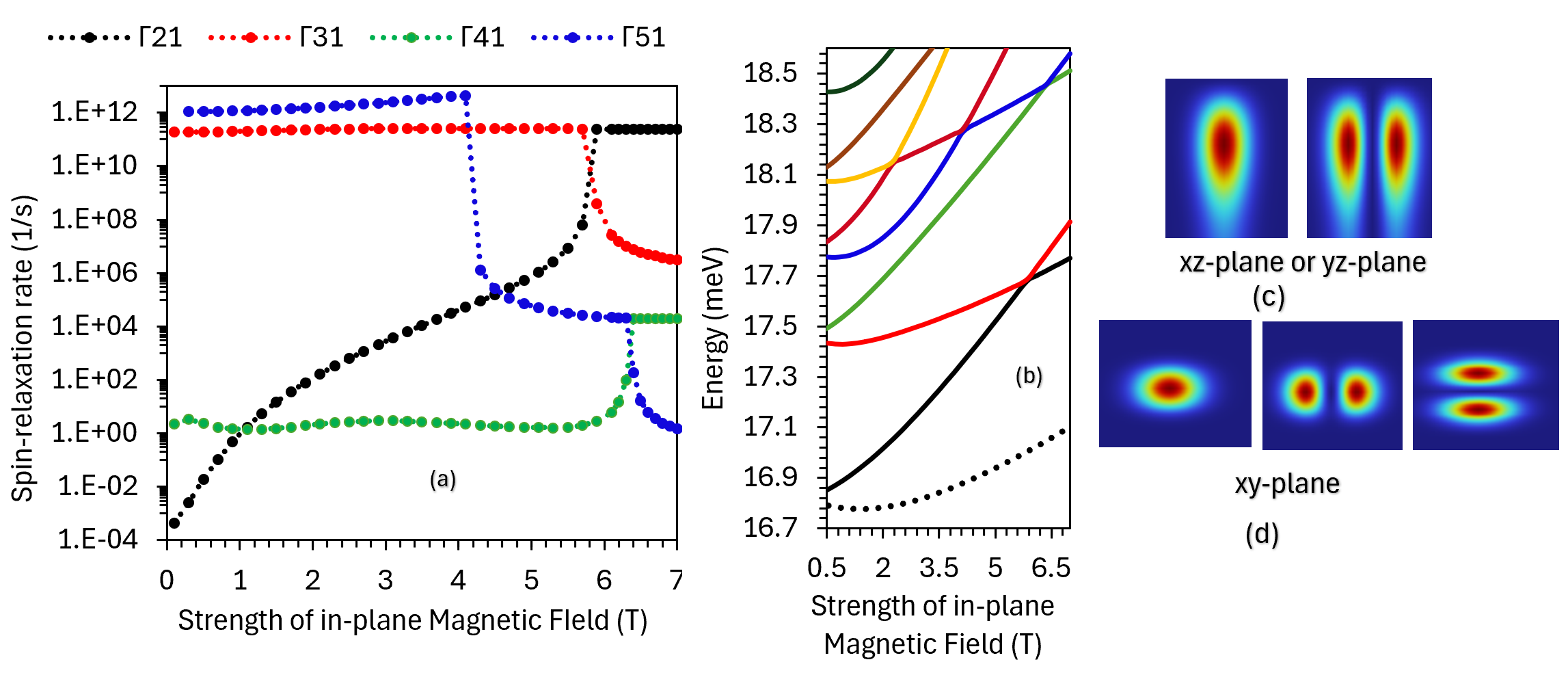}
\caption{
\label{fig3} (Color online) \hc{(a) Rates $\Gamma N1$ versus strength of in-plane field $B_{\mbox{\scriptsize in}}$ in a double QD of $\ell_0$ = 25 nm separated at $y_0$ = 10 nm from the origin. Two hot spots emerge at 5.9 and 6.4 T due to the avoided crossing between the bands shown in (b).  Probability densities of the ground and the 2nd excited state in $xz$ or $yz$ plane are illustrated in (c), while those of the ground, 2nd excited and 4th excited state in $xy$ plane are in (d). We chose $B_z$ = 0.2 T. In (c) and (d), we also chose $B_{\mbox{\scriptsize in}}$ = 0.2 T.}
}.
\end{figure*}

\hc{\subsection{Variation of magnetic field strength}}

\hc{The four spin-relaxation rates versus $B_{\mbox{\scriptsize in}}$ in the presence of a weak $B_z$ = 0.2 T for a double QD displaced at $\pm$10 nm from the origin or 20 nm inter-dot distance are displayed in Fig.~\ref{fig3}. Note that the rate $\Gamma 41$ is found very weak and rather insensitive to $B_{\mbox{\scriptsize in}}$ over a large range until near the hot spot. More importantly, two hot spots appear, one between $\Gamma 21$ (black) and $\Gamma 31$ (red) and another between $\Gamma 41$ (green) and $\Gamma 51$ (blue). Remarkably again, the rate of the latter is several orders of magnitude smaller than that of the former, resulting in a huge relaxation time increase. In specific, the $\Gamma$ value of the latter hot spot being around $10^3$ s$^{-1}$ yields a decoherence time ($T_2 \le 2T_1$) in the millisecond range. We comment that these very long-lived hot-spot activities may provide a suitable mechanism for spin energy transfer between the two QDs for qubit gate operations. The band structures of the double QD are shown in Fig.~\ref{fig3}(b). The avoided level-crossings seen involving the bands at around $B_{\mbox{\scriptsize in}}$ = 5.9 T and 6.4 T are clearly responsible for inducing hot spots in Fig.~\ref{fig3}(a)}. The probability densities of the spin and orbital states are shown in Fig.~\ref{fig3}(c) and (d). Note that the densities associated with spin splitting and orbital splitting due to the applied magnetic fields are not shown in this figure, but they are identical to those shown in Fig.~\ref{fig3}(c) and (d). Finite-size effects are observed in the probability density of the orbital states, as seen in the $xy$ plane of the two right panels of Fig.~\ref{fig3}(d). Furthermore, the spin-relaxation behavior in Fig.~\ref{fig3} is similar to that in Fig.~\ref{fig1}. This is because a double-dot separation of 10 nm from the origin is not large enough to distinguish between a single and a double QD. We performed calculations for several out-of-plane magnetic field values, including $B_z$ = 0 T, and found no significant qualitative difference from Fig.~\ref{fig3}.
\begin{figure*}
\includegraphics[width=15.0cm,height=8cm]{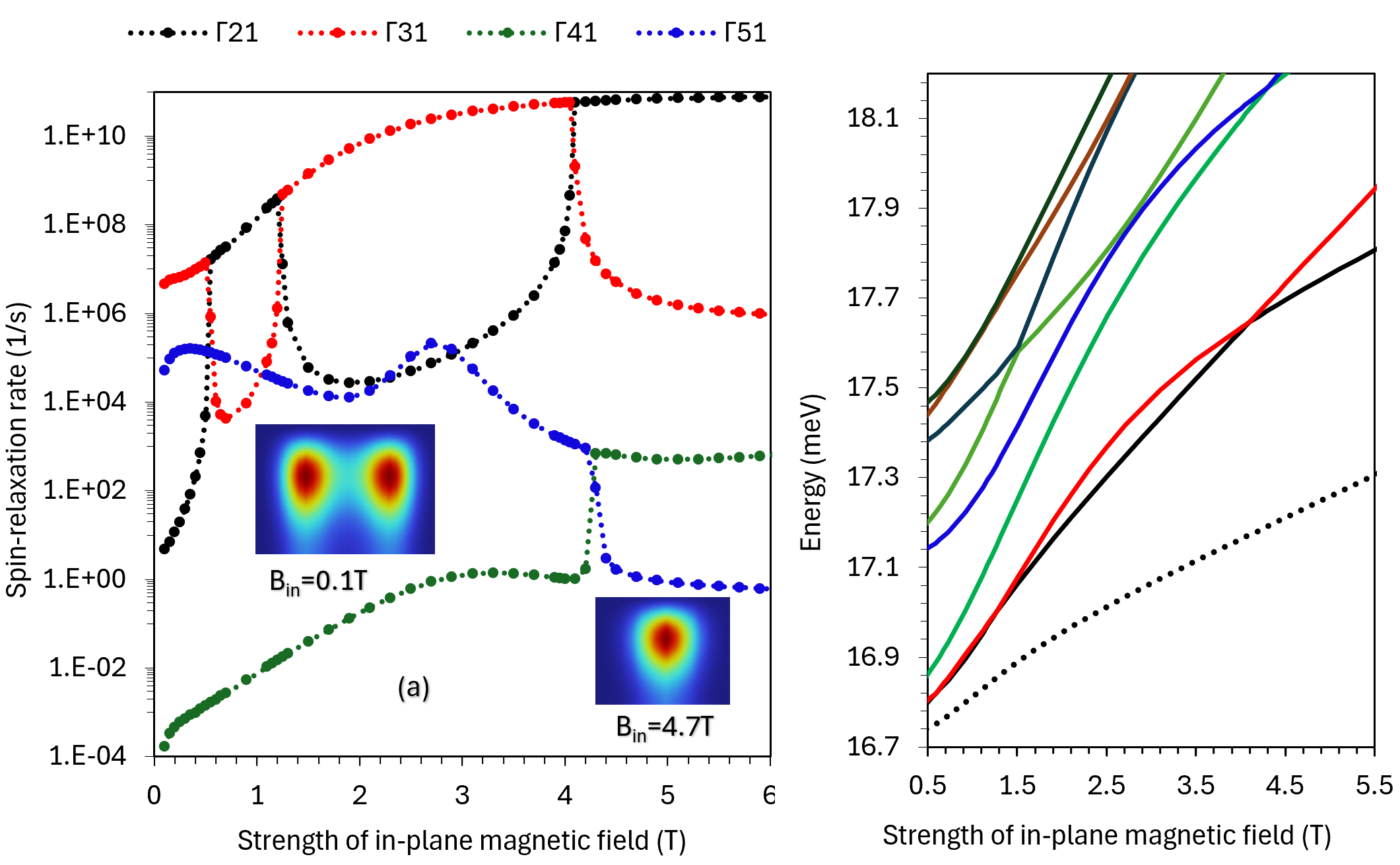}
\caption{
\label{fig4} (Color online) \hc{(a) Rates $\Gamma N1$ versus strength of in-plane field $B_{\mbox{\scriptsize in}}$ in a double QD of $\ell_0$ = 25 nm separated at $y_0$ = 60 nm from the origin. In addition to the two spin-hot spots at 4.1 T and 4.3 T, two additional hot spots at lower 0.5 and 1.2 T appear. The corresponding avoided-crossings of the bands can be seen in (b).  Probability densities of ground states at different values of $B_{\mbox{\scriptsize in}}$, noted in the figure, are shown in the insets. We chose $B_z$ = 0.2 T.}
}.
\end{figure*}

\hc{Fig.~\ref{fig4}(a) illustrates the spin-relaxation rates for a double QD displaced by $y_0$  = 60 nm from their centers, that is for an inter-dot separation of 120 nm, as $B_{\mbox{\scriptsize in}}$ varies. We still keep $B_z$ = 0.2 T. The spectra produce two hot spots, as in Fig.~\ref{fig3}(a), at higher values of $B_{\mbox{\scriptsize in}}$ in the vicinity of 4.0 T. Among these, the hot spot shown by the green and blue curves, due to the interaction of $\Gamma 41$ and $\Gamma 51$, features, as previously, about six orders of magnitude smaller rate than the other by $\Gamma 21$ (black) and $\Gamma 31$ (red); the corresponding avoided crossings can be seen in panel (b). Following the identical trend as in Fig.~\ref{fig3}(a), the former spot with milliseconds resonant lifetime can qualify as a candidate for quantum entanglement and information processing. Since this double QD has an inter-dot separation of 120 nm, we may expect the formation of two distinct dots in the probability density image. However, while $B_{\mbox{\scriptsize in}}$ = 0.1 T meets this expectation, we observe a single dot in the bottom inset of the probability density plot at $B_{\mbox{\scriptsize in}}$ = 4.7 T. This is because at this higher $B_{\mbox{\scriptsize in}}$ sector the resulting magnetic confinement potential overpowers the gate potential ($\omega_c/\omega_0 \approx 4.2$) to induce an effective one dot density. In any event, at smaller fields ($B_{\mbox{\scriptsize in}}$ = 1.0 and 1.5 T) however, two additional hot spots, in the microsecond relaxation-time regime, show up in Fig.~\ref{fig4}(a) between the rates $\Gamma 21$ (black) and $\Gamma 31$ (red). Correspondingly, the respective bands induce additional avoided-crossings at these values as seen in Fig.\ref{fig4} (b). Such unusual  band interactions at lower-$B_{\mbox{\scriptsize in}}$ can be attributed to the stronger effect on the lowest bands, than on other bands, of the non-parabolicity term. This term originates from the $k \cdot p$ theory in the simplest 2-band calculations due to the inclusion of spin-orbit couplings, particularly when the confinement due to the gate potential is stronger than the magnetic potential~\cite{sp13jap}. We clearly observe the formation of a double dot density structure at small magnetic fields, as shown in the upper inset of the probability density plot for ($B_{\mbox{\scriptsize in}}$ = 0.1 T). In this specific instance, the confinement due to the gate potential is indeed stronger than the magnetic confinement induced by combined $B_{\mbox{\scriptsize in}}$ = 0.1 T and $B_z$ = 0.2 T fields ($\omega_c/\omega_0 \approx 0.1$).}
\begin{figure*}
\includegraphics[width=17cm,height=8cm]{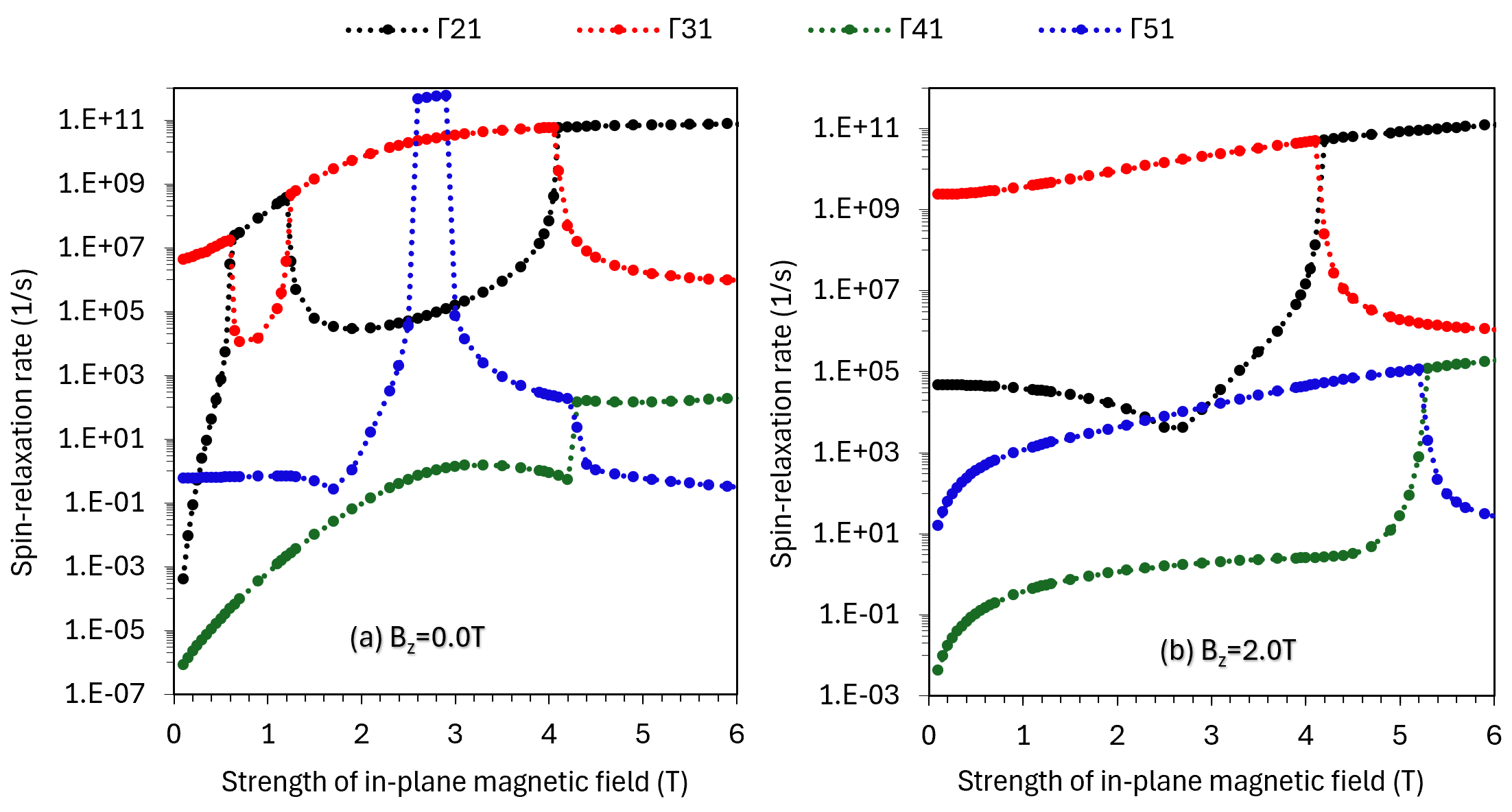}
\caption{
\label{fig5} (Color online) \hc{Spin-relaxation rates versus $B_{\mbox{\scriptsize in}}$ in a double QD separated at $y_0$ = 60 nm from the origin comparing $B_z$ = 0 T (a) with $B_z$ = 2 T (b).}
}
\end{figure*}

Fig.~\ref{fig5} compares the spin-relaxation rates as a function of $B_{\mbox{\scriptsize in}}$ in a double QD at $y_0$ = 60 nm between $B_z$ = 0 T (a) with $B_z$ = 2 T (b), respectively. Plots in panel (a) are qualitatively similar to Fig.~\ref{fig4}(a): Here, even in the absence of $B_z$, we also observe additional hot spots at lower $B_{\mbox{\scriptsize in}}$ = 0.5 and 1.2 T emanating from the $\Gamma_{21}$ and $\Gamma_{31}$ interaction. This suggests that when the dot separation is large, and the magnetic confinement potential is weak, oscillations in the band structures (as also seen Fig.~\ref{fig4}(b)) can induce such lower $B$-field spin-hot spots with somewhat weakened relaxation rates slowed down to microseconds. \hc{With a large enough $B_z$, however, even for a double QD with large separation a stronger vector potential ($\mathbf{A}$) leads to in-plane circulation $\mathbf{\nabla} X \mathbf{A}$, which can dominate. This may effectively restore hot-spot properties of a double QD of shorter separation, such that the low $B_{\mbox{\scriptsize in}}$ spots will disappear, while spots at higher $B_{\mbox{\scriptsize in}}$ survive. This is why Fig.~\ref{fig5}(b) with a strong $B_z$ but longer separation looks similar to Fig.~\ref{fig3}(a) of weak $B_z$ yet shorter separation.}

\hc{\subsection{Useful insights}}

Beyond the mathematical description provided by Eqs.~(\ref{had-2})--\ref{H2-2}), the emergence of additional spin-hot spots can be understood qualitatively from the evolution of the electronic spectrum as the dots are spatially separated. In a silicon double QD, acoustic-phonon-mediated spin relaxation is enabled by spin–orbital mixing induced by spin–orbit coupling, and the corresponding relaxation rate depends sensitively on the energy separation and spatial character of the participating states. As the inter-dot separation increases, \hc{the strength of tunnel-coupling falls almost exponentially with the separation. More precisely,} the overlap between the left- and right-dot wave functions decreases, reducing the effective coupling between the dots and modifying the bonding–antibonding and orbital level splittings. As a consequence, orbital and spin-dependent levels can approach one another and undergo additional crossings or avoided crossings, where spin–orbit coupling enhances the mixing between states of different spin character. Each such avoided crossing provides a potential spin-hot spot, leading to multiple relaxation features as a function of inter-dot separation and magnetic field. \hc{Thus, a relaxation map plotted against inter-dot displacement and magnetic field would likely exhibit multiple ridges of enhanced relaxation. This is consistent with the general trend observed in parts of our results.} Particularly, our calculations show that when the dots are displaced by approximately 60 nm, the competition between gate-induced confinement and magnetic confinement produces additional level crossings and corresponding low- and high-field spin-hot spots. These results demonstrate that spatial displacement \hc{can provide a sensitive control parameter} for engineering spin relaxation in silicon double QDs.


\vspace{0.3 cm}
\section{Possible experimental realization}\label{experimentalA}

\hc{The predicted separation-induced hot spots can be tested in gate-defined silicon double QDs (Si MOS or Si/SiGe) of the kind already used for spin-relaxation and spin-readout studies~\cite{jock18,lundberg20,ciriano21,hao14}. In such devices, plunger and barrier gates set the effective dot separation and tunnel coupling~\cite{zhao22}, while a vector magnet sets $B_{\rm in}$ and $B_z$ independently. We propose a two-step test. First, microwave (ESR/EDSR) or photon-assisted-tunneling spectroscopy~\cite{hao14} versus $B_{\rm in}$ at fixed gate settings should resolve the avoided crossings of Figs.~\ref{fig3}(b) and \ref{fig4}(b) directly, fixing the level structure that drives the hot spots. Second, single-shot spin-to-charge readout based on spin-selective tunneling or Pauli spin blockade~\cite{lundberg20,ciriano21,takeda24}, applied after pulsed initialization of the excited level, gives $1/T_1$ versus $B_{\rm in}$, with peaks expected at the predicted fields (e.g., near $4.1$ and $4.3~\mathrm{T}$ for $y_0=60~\mathrm{nm}$, Fig.~\ref{fig4}(a)). Millisecond lifetimes ($\Gamma\sim10^{3}~\mathrm{s^{-1}}$) lie well within the reach of single-shot methods. Microsecond hot spots require fast reflectometry readout, while still faster ones would appear as a collapse of readout contrast. Three signatures separate the proposed mechanism from other relaxation channels: (i) the hot-spot field shifts with the gate-tuned separation along the locus of Fig.~\ref{fig2b}(c); (ii) no hot spots occur for $1.8\lesssim B_{\rm in}\lesssim3.8~\mathrm{T}$; and (iii) raising $B_z$ to $\approx2~\mathrm{T}$ removes the low-field hot spots while the high-field ones persist (Fig.~\ref{fig5}).}

\hc{The experimental requirements are modest. A tilt $\theta$ produces $B_z=B_{\rm in}\sin\theta$, so alignment to $1^\circ$ keeps $B_z<0.1~\mathrm{T}$ at $5~\mathrm{T}$, far below the $2~\mathrm{T}$ close to which the spectrum changes (Fig.~\ref{fig5}). Dilution-refrigerator temperatures lie far below $E_Z/k_B\approx5$--$7~\mathrm{K}$ at $4$--$5.5~\mathrm{T}$, consistent with the phonon-emission-only regime of Eq.~\eqref{1-T1}. The main confounding mechanism is spin-valley mixing, which our single-valley effective-mass model omits. It produces a hot spot where the Zeeman energy equals the valley splitting~\cite{Yang2013}, and typical gate-tunable valley splittings of $0.3$--$0.8~\mathrm{meV}$ overlap the Zeeman energies of $\approx0.46$--$0.64~\mathrm{meV}$ ($g=2$) at $4$--$5.5~\mathrm{T}$. Because the separation-induced hot spots follow the gate-defined separation $2y_0$ while valley hot spots are set by the vertical field, sweeping $2y_0$ at fixed vertical field, together with an independent measurement of the valley splitting, can distinguish the two. Such control of spin hot spots may provide a useful route for engineering spin relaxation and coherence in silicon-based quantum-information devices.}

\vspace{0.75 cm}
\section{Conclusion}\label{conclusion}

\hc{We have studied acoustic-phonon-mediated spin relaxation in single and laterally displaced silicon double quantum dots under combined in-plane ($B_{\rm in}$) and out-of-plane ($B_z$) magnetic fields, using finite-element diagonalization of a spin-orbit-coupled effective-mass Hamiltonian. For a single dot, the relaxation rate depends strongly on $B_z$ only below $B_{\rm in}\approx1.5~\mathrm{T}$ and shows a hot spot near $5.5~\mathrm{T}$ (Fig.~\ref{fig1}). Displacing the dots produces additional, non-adiabatic hot spots that a single dot does not have. At $B_z=B_{\rm in}=0.2~\mathrm{T}$, an avoided crossing between the second and third excited levels gives a separation-induced hot spot at a center-to-center distance of about $130~\mathrm{nm}$ (Fig.~\ref{fig2}). The quadratic dependence of the double-dot Hamiltonian on the separation, Eqs.\,(\ref{had-2}) and (\ref{H2-2}), accounts for the parabolic level crossings behind it. At larger in-plane fields, new hot spots appear from the avoided crossing of the fourth and fifth levels, with rates many orders of magnitude lower, and no hot spots occur for $1.8\lesssim B_{\rm in}\lesssim3.8~\mathrm{T}$ (Fig.~\ref{fig2b}).}

\hc{Sweeping $B_{\rm in}$ at fixed separation shows how gate and magnetic confinement compete. For nearby dots ($y_0=10~\mathrm{nm}$), the response resembles that of a single dot (Fig.~\ref{fig3}). For well-separated dots ($y_0=60~\mathrm{nm}$), two high-field hot spots near $4.1$ and $4.3~\mathrm{T}$ are accompanied by two low-field hot spots at $0.5$ and $1.2~\mathrm{T}$ (Fig.~\ref{fig4}). The high-field hot spot formed by the fourth and fifth levels has a relaxation rate about six orders of magnitude below that of the other, reaching the millisecond regime ($\Gamma\sim10^{3}~\mathrm{s^{-1}}$ for $y_0=10$ as well as $60~\mathrm{nm}$), whereas the second and third levels relax far faster. A strong out-of-plane field ($B_z=2~\mathrm{T}$) removes the low-field hot spots and keeps the high-field ones (Fig.~\ref{fig5}). Displacement of the dots is therefore a control parameter for placing spin-relaxation resonances and selecting long-lived operating points, which is relevant for spin-qubit gate operations. Our model treats a single valley and phonon emission only, and it neglects charge noise and hyperfine interaction. The proposed measurements in Sec.~V, together with an independent determination of the valley splitting, would provide a direct test. Extending the model to include valley physics can be a natural next step.}

~\\

\section{Acknowledgements}\label{acknowlwdgement}
\hc{The simulations were performed at BARTIK High-Performance cluster, supported by the U.S. National Science Foundation Grant No. CNS-1624416, at Northwest Missouri State University.} S.P. acknowledges support from the School of Natural Science, Northwest Missouri State University for purchasing COMSOL multiscale multiphysics simulations software package. This study also acknowledges the Visiting Faculty Program (2026), sponsored by the U.S. Department of Energy and hosted by Stanford University and SLAC National Accelerator Laboratory. \hc{H.S.C. acknowledges supports from the U.S. National Science Foundation Grant No. PHY-2512850.} R.M. is acknowledging the support of NSERC Discovery and CRC Programs.


%

\end{document}